\documentclass[10pt,aps,prd,twocolumn,superscriptaddress,nofootinbib]{revtex4-1}
\usepackage[usenames]{xcolor}
\usepackage{amsmath,amssymb}
\usepackage{mathrsfs}
\usepackage{amsthm}
\usepackage{hyperref}
\usepackage{appendix}
\newcommand{\ie}{\emph{i.e.}}
\newcommand{\eg}{\emph{e.g.}}
\newcommand{\threeR}{\mathscr{R}}%{{}^{(3)}{R}}
\newcommand{\met}{\mathsf{g}}
\newcommand{\dd}{\mathrm{d}}
\newcommand{\aS}{a_{\textsc{s}}}
\newcommand{\kS}{k_{\textsc{s}}}
\newcommand{\cg}{c_{\gamma}}
\newcommand{\wm}{w_{\mathrm{m}}}
\newcommand{\order}{\boldsymbol{\mathcal{O}}}
\begin{document}
%======================================================================================================================
\title{de Sitterization via Kerr-Schild}
%======================================================================================================================
\author{Aditya Sharma}
\email{p20170442@goa.bits-pilani.ac.in}
\affiliation{BITS-Pilani, KK Birla Goa Campus, NH 17B, Bypass Road, Zuarinagar, Goa, India 403726}

\author{Kinjal Banerjee}
\email{kinjalb@gmail.com}
\affiliation{BITS-Pilani, KK Birla Goa Campus, NH 17B, Bypass Road, Zuarinagar, Goa, India 403726}

\author{Jishnu Bhattacharyya}
\email{jishnub@gmail.com}
%\affiliation{}
%======================================================================================================================
%\date{May(?) 2022; revised July 2022}
%======================================================================================================================
\begin{abstract}
Spatially homogeneous cosmological spacetimes, evolving in the presence of a positive cosmological constant and matter satisfying some reasonable energy conditions, typically approach the de Sitter geometry asymptotically (at least locally). In this work, we propose an alternate way to characterize this phenomena. We focus on a subset of such models admitting a generalized Kerr-Schild representation. We argue that the functions which define such a representation can be chosen, such that, their asymptotic behavior make the evolution towards the de Sitter spacetime manifest through the representation. We verify our claim for the Kantowski-Sachs family of cosmological spacetimes.
\end{abstract}
%======================================================================================================================
\maketitle
%======================================================================================================================
\section{Introduction}
It is well known that a positive cosmological constant is the best `isotropizer'~\cite{Starobinsky:1982mr}, in the sense that expanding cosmological models in the presence of a positive cosmological constant typically approach the de Sitter spacetime asymptotically~\cite{Gibbons:1977mu}. Such statements can be turned into a precise theorem for a large class of spatially homogeneous cosmological models. This result, to be referred to as~\emph{Wald's theorem}~\cite{Wald:1983ky} henceforth, can be stated as follows:
        \begin{itemize}
         \item an initially expanding solution of a homogeneous cosmological model,
         \item characterized by a non-positive scalar (Ricci) curvature $\threeR$ of the surfaces of homogeneity (\ie, $\threeR \leqq 0$),
         \item evolving under the influence of a positive cosmological constant (\ie, $\Lambda > 0$),
         \item and matter satisfying the dominant and the strong energy conditions
        \end{itemize}
must continue to expand indefinitely, isotropize, and be locally indistinguishable from the de Sitter spacetime asymptotically. For reasons to be clarified below, we will call this process~\emph{de Sitterization}.

Our presentation of Wald's theorem differs from the original formulation, in that, in~\cite{Wald:1983ky} the theorem only concerned Bianchi type cosmological models and it was actually~\emph{proven} that all Bianchi models except Bianchi IX respect the condition $\threeR \leqq 0$. However, even in models like Bianchi IX where this condition may not hold, de Sitterization may still take place under suitable additional conditions. Keeping this condition as a separate assumption therefore helps us to understand the precise role of this particular condition in the de Sitterization process.

Two remarkable aspects of Wald's theorem are the generality of its premise and the simplicity of the underlying arguments. Indeed, as long as the aforementioned conditions are satisfied, the result is neither sensitive to the detailed nature of the matter present nor to most of the dynamical equations governing the overall evolution of the spacetime and matter. We present a brief outline of the proof of the theorem in~\ref{appendix:WaldsTheoremProof}.

On the other hand, the generality of the arguments also makes it difficult to explicitly understand how the result may (or may not) fail when one of the underlying assumptions is violated. In fact, the various FLRW models offer an excellent illustration of this point. Consider, for example, the evolution of the FLRW models in the presence of a positive cosmological constant and a perfect fluid satisfying both the dominant and the strong energy conditions. Then, every initially expanding solution of both the flat and open FLRW models de Sitterizes, as dictated by Wald's theorem. On the other hand, the situation is more interesting with the closed FLRW models. Here, some initially expanding solutions do de Sitterize, but others generically recollapse and end up in a `big crunch'.

Incidentally, these FLRW examples justify our coinage of the term `de Sitterization'. Every FLRW spacetime is isotropic and so it is meaningless to talk about their isotropization, but a FLRW spacetime may or may not de Sitterize as we just noted. In other words, the term `de Sitterization' can unambiguously be applied to both isotropic and initially anisotropic spacetimes evolving towards the de Sitter spacetime\footnote{We should also note in this regard that the term `isotropization' can describe more general situations that `de Sitterization'. For instance, an anisotropic Bianchi I spacetime may evolve into a non-de Sitter FLRW spacetime in the absence of any cosmological constant. This is an example of isotropization, but not of de Sitterization.}.

In order to understand the issues raised in the preceding paragraphs more closely, it is customary to employ the techniques of dynamical systems analysis. In this approach, we exploit the fact that the evolution equations of a homogeneous cosmological model form a set of coupled first order ordinary differential equations, and therefore, can be viewed as describing a dynamical system. This is well established research program (see~\cite{Ellis:1998ct},~\cite{EW:1997},~\cite{Coley:2003mj},~\cite{Coley:1999uh},~\cite{Goliath:1998na} and references therein) which has made significant progress in describing various cosmological models with different types of matter content in the context of General Relativity and in extended theories of gravity \cite{Boehmer:2014vea}.

Detailed analyses of the homogeneous cosmological models within this framework  reveal that forgoing the $\threeR \leqq 0$ assumption typically leads to the existence of additional fixed points which are usually saddle points. The resulting state space gets partitioned, such that some orbits describe de Sitterization while others describe recollapse, and the boundaries between such regions consist of metastable states. One minor drawback of the dynamical systems approach is that one might need to redo the analysis if the matter content of the model is modified. However, the process always yields much more detailed information about the dynamics of the model.
%----------------------------------------------------------------------------------------------------------------------
\subsection*{\textbf{Our proposal:}}
In the present work, we wish to describe the de Sitterization process in yet another way. Our proposal can be stated as follows: let $\met_{a b}$ be the physical metric on a de Sitterizing cosmological spacetime, and suppose that it can be expressed in a~\emph{generalized Kerr-Schild form}~\cite{Stephani:2003tm} as follows
        \begin{equation}\label{KerrSchild:met}
         \met_{a b} = \Omega^{2}\tilde{\met}_{a b} - 2\Phi\ell_{a}\ell_{b}~,
        \end{equation}
where $\tilde{\met}_{a b}$ is the `de Sitter metric' (more precisely, a conformally flat Einstein metric whose curvature is determined by the physical cosmological constant $\Lambda > 0$), $\ell_{a}$ is a real null one-form of the physical metric, and $\Omega$ and $\Phi$ are a pair of functions. Then, we claim that~\emph{for appropriate choices for the functions $\Omega$ and $\Phi$ and the null one-form $\ell_{a}$, the above relation captures the de Sitterization process}.

Before proceeding further some clarifications of our claim seem warranted. Perhaps it is best to start with the choice of the two functions and the null one-form appearing in equation~\eqref{KerrSchild:met}. Suppose for a moment that $\ell_{a}$ is an arbitrary real null one-form of the physical metric, and likewise, let $\Phi$ be an arbitrary function on the spacetime. It is then very easy to show that as long as the function $\Phi$ and the one-form $\ell_{a}$ are non-singular, the tensor $\hat{\met}_{a b} \equiv \met_{a b} + 2\Phi\ell_{a}\ell_{b}$ is a symmetric rank-$(0, 2)$ non-degenerate tensor with the same signature as that of the physical metric. In other words, $\hat{\met}_{a b}$ may define~\emph{some} metric on the spacetime. However, for arbitrary choices of $\Phi$ and $\ell_{a}$ one can hardly expect $\hat{\met}_{a b}$ to have any physical relevance. In other words, for the relation~\eqref{KerrSchild:met} to have any physical significance whatsoever, one has to carefully prescribe the properties of the quantities appearing on the right hand side. Note that the function $\Omega$ can never vanish since it would otherwise make the conformally de Sitter metric $\Omega^{2}\tilde{\met}_{a b}$ degenerate (which can never happen as long as $\Phi$ is non-singular). Hence, $\Omega$ can never change sign and we can always assume $\Omega > 0$.

One reasonable assumption, then, is to impose all the Killing symmetries of the physical metric on the functions $\Omega$ and $\Phi$ (since we do want these functions to contain physical information about the evolution of the solutions). Furthermore, a prerequisite to the metric $\tilde{\met}_{a b}$ being de Sitter is that it be conformally flat. This last condition usually puts significant restrictions on the physical metric, because it will not likely hold unless the physical metric is algebraically special. In fact, under such conditions, the null vector $\ell^{a} = \met^{a b}\ell_{b}$ must become a principal null vector of the conformal (Weyl) curvature of the physical metric~\cite{Stephani:2003tm}, and therefore, will also satisfy all of the Killing symmetries. One may thus deduce that the metric $\tilde{\met}_{a b}$ must admit all the Killing symmetries of the physical metric as well\footnote{Of course, the de Sitter metric, being maximally symmetric, will admit more Killing symmetries than the physical metric. Our assumptions merely imply that a subset of these symmetries will be shared with the physical metric.}.~\emph{The primary goal of this paper is to demonstrate, through a specific example, that the above requirements are enough to determine the functions $\Omega$ and $\Phi$ uniquely, and more importantly, that these functions have the right kind of asymptotic behavior to capture de Sitterization.}

An obvious shortcoming of the proposed approach is that it is only possible to study algebraically special solutions (typically Petrov type D, at most type II) in this manner. However, whenever applicable, this approach can be quite useful and is intended to complement the well established approaches to understand de Sitterization. Indeed, the seemingly  `perturbative appearance' of the relation~\eqref{KerrSchild:met} -- where one expresses the physical metric as its asymptotic de Sitter form (up to a conformal factor) plus a `correction term' (the $-2\Phi\ell_{a}\ell_{b}$ piece) -- makes it rather easy to visualize the de Sitterization process. When chosen correctly, the function $\Omega$ is expected to approach unity asymptotically for a de Sitterizing solution, while $\Phi$ is expected to vanish in the same limit, thereby forcing the physical metric to approach its asymptotic de Sitter form. We will demonstrate this behavior explicitly for our example below. Such behavior thus offers us a~\emph{coordinate independent} way to capture useful information about asymptotic behavior of such de Sitterizing spacetimes through the functions $\Omega$ and $\Phi$. However, despite its `perturbative appearance', the relation~\eqref{KerrSchild:met} is actually fully non-perturbative, and therefore describes a de Sitterizing solution far away from the de Sitter fixed point. We will elaborate on these (and other) points further in our concluding remarks in section~\ref{section:conclusion}.

The rest of this paper is devoted to the analysis of a specific example in order to substantiate our proposal. This example, the~\emph{Kanstowski-Sachs family of spacetimes}, is introduced in section~\ref{section:KS}, and properties of such spacetimes relevant for discussing de Sitterization are reviewed. The generalized Kerr-Schild decomposition of these spacetimes,~\emph{\`a la}~\eqref{KerrSchild:met} is presented in section~\ref{section:KSKS}, and the demonstration of the required asymptotic behavior of the functions $\Omega$ and $\Phi$ for de Sitterizing solutions is carried out in section~\ref{section:KSKS:soln}. We end with a discussion of our results in section~\ref{section:conclusion}. We have also included two appendices:~\ref{appendix:WaldsTheoremProof} contains an outline of the proof of Wald's theorem, and~\ref{appendix:FLRW} discusses a conformal representation of de Sitterizing FLRW solutions analogous to~\eqref{KerrSchild:met}.
%======================================================================================================================
\section{The Kantowski-Sachs family of spacetimes}\label{section:KS}
As mentioned in our introductory remarks, we wish to illustrate our proposal~\eqref{KerrSchild:met} through the example of the~\emph{Kantowski-Sachs family of spacetimes}\footnote{The terminology is borrowed from reference~\cite{Maccallum:1980gd}.}. We can define them as a family of homogeneous spacetimes whose symmetries constrain the metric to take the following form
        \begin{equation}\label{KaSa:met}
         \dd{s}^2 = -\dd{t}^{2} + a_{z}^{2}\dd{z}^{2} + a_{\gamma}^{2}(\dd\theta^{2} + S_{\gamma}(\theta)^{2}\dd\varphi^{2})~.
        \end{equation}
Here, the time coordinate $t$ parametrizes the homogeneous hypersurfaces and the metric components $a_{z}$ and $a_{\gamma}$ are functions of $t$ only. The coordinates $\theta$ and $\varphi$ on the~\emph{transverse space} need not be compact (here and henceforth, any function carrying the subscript $\gamma$ will pertain to the transverse space, unless specified otherwise), and the function $S_{\gamma}(\theta)$ can take one of the three possible forms, namely: $\sinh\theta$, $\theta$, or $\sin\theta$.

The scalar curvature $\threeR$ of the homogeneous hypersurfaces is given by
        \begin{equation}\label{KaSa:3R}
         \threeR = \frac{2\cg}{a_{\gamma}^{2}}~,
        \end{equation}
where the constant $\cg$ is defined via the relation $\cg = -S''_{\gamma}(\theta)/S_{\gamma}(\theta)$. Hence,
        \begin{equation*}
         \cg =
          \begin{cases}
           -1~, & \qquad~\text{for}~S_{\gamma}(\theta) = \sinh\theta~, \\
            0~, & \qquad~\text{for}~S_{\gamma}(\theta) = \theta~, \\
            1~, & \qquad~\text{for}~S_{\gamma}(\theta) = \sin\theta~.
         \end{cases}
        \end{equation*}
In other words, the choice of the function $S_{\gamma}(\theta)$ determines the curvature of the homogeneous hypersurfaces.

The metric~\eqref{KaSa:met} admits two obvious Killing vectors, namely $\partial_{z}$ and $\partial_{\varphi}$, the former being orthogonal to the transverse space, and the latter being inside it. There are also two additional Killing vectors, both residing in the transverse space, given by
        \begin{equation*}
         \sin\varphi\,\partial_{\theta} + [S'_{\gamma}(\theta)/S_{\gamma}(\theta)]\cos\varphi\,\partial_{\varphi}~,
        \end{equation*}
and
        \begin{equation*}
         \cos\varphi\,\partial_{\theta} - [S'_{\gamma}(\theta)/S_{\gamma}(\theta)]\sin\varphi\,\partial_{\varphi}~.
        \end{equation*} 
Together, these four Killing vectors generate the Lie algebra of the $\mathcal{G}_{4}$ symmetry group of such spacetimes. In each case the parent $\mathcal{G}_{4}$ admits a $\mathcal{G}_{3}$ subgroup, and in all but one cases the symmetry acts in a simply-transitive manner. All such cases are~\emph{locally rotationally symmetric (LRS)} Bianchi models; in particular, for $S_{\gamma}(\theta) = \sinh\theta$ the model is LRS Bianchi III, while for $S_{\gamma}(\theta) = \theta$ it is LRS Bianchi I. The only exception occurs with the subfamily of spacetimes with $S_{\gamma}(\theta) = \sin\theta$ where the symmetry group acts multiply-transitively. These are the proper Kantowski-Sachs models, introduced in~\cite{Kantowski:1966te} and~\cite{Kantowski:thesis}. For further information about these models, see~\cite{Maccallum:1980gd},~\cite{Ellis:1998ct} and references therein.

In order to describe the de Sitterization process efficiently, it is useful to introduce the~\emph{scale factor} $a$, given in terms of the metric functions $a_{z}$ and $a_{\gamma}$ as 
        \begin{equation}\label{def:a}
         a^{3} = \frac{a_{z}a_{\gamma}^{2}}{\tau_{0}^{2}}~,
        \end{equation}
where $\tau_{0}$ is a (time) scale associated with the (positive) cosmological constant $\Lambda$ according to
        \begin{equation}\label{def:tau0}
         \tau_{0} = \sqrt{\frac{3}{\Lambda}}~.
        \end{equation}
The time derivative of the scale factor then allows us to define the Hubble parameter $H$, as usual. However, instead of working with the conventionally defined $H$, it turns out to be more convenient to introduce a~\emph{dimensionless Hubble parameter} $h$ as follow   
        \begin{equation}\label{EOM:a}
         h = \frac{\tau_{0}}{a}\frac{\dd{a}}{\dd{t}}~.
        \end{equation}
The conventional Hubble parameter may then be related to $h$ through $H = \tau_{0}^{-1}h$. In particular, in a de Sitterizing solution, $H$ approaches the value $\tau_{0}^{-1}$ asymptotically; hence, $h$ approaches unity in the same limit.

We may also define a~\emph{relative scale factor} $\aS$ through the following relation
        \begin{equation}\label{def:aS}
         \aS = \frac{\tau_{0}a_{z}}{a_{\gamma}}~.
        \end{equation}
The metric functions $a_{z}$ and $a_{\gamma}$ both diverge as the de Sitter limit is approached. However, they diverge at the same rate such that the relative scale factor tends to unity asymptotically.  Moreover, the time derivative of $\aS$ allows us to introduce the function $\kS$ as below 
        \begin{equation}\label{EOM:aS}
         \kS = \frac{\tau_{0}}{\aS}\frac{\dd\aS}{\dd{t}}~,
        \end{equation}
which has the following interpretation: as is well known, the trace-free part of the extrinsic curvature of the homogeneous hypersurfaces provides a measure of anisotropy of homogeneous cosmological models. In the Kantowski-Sachs family of spacetimes, in particular, the symmetries dictate that the said trace-free part can be described by a single function of $t$. The function $\kS$ introduced above, is  essentially the dimensionless part of that function. Naturally, in a de Sitterizing solution, $\kS$ is expected to vanish asymptotically. To summarize, what we have done so far is to capture the degrees of freedom of the metric~\eqref{KaSa:met} and their time derivatives in the variables $a$, $\aS$, $h$ and $\kS$. Apart from their nice geometrical interpretations, these variables are useful in expressing some of our results in succinct and more illuminating form.

To consider dynamics, in addition to the cosmological constant $\Lambda$, we need to specify the matter stress tensor to source the Einstein's equations. We take this to be that of a perfect fluid whose flow lines are perpendicular to the homogeneous hypersurfaces (as dictated by the symmetries) and whose pressure $p$ and energy density $\rho$ are linked by an equation of state of the form
        \begin{equation}\label{KaSa:eos}
         p = \wm\rho~.
        \end{equation}
In accordance with the assumptions behind Wald's theorem, we require the fluid to respect both the strong and dominant energy conditions. These restrict the constant $\wm$ to be bounded from both above and below according to 
        \begin{equation}\label{wm:SEC+DEC}
         -\frac{1}{3} \leqq \wm \leqq 1~.
        \end{equation}
The stress tensor conservation equation now allows the energy density to be related to the scale factor through
        \begin{equation}\label{rho(a)}
         \rho = \frac{m_{0}\tau_{0}^{-2}}{a^{3(1 + \wm)}}~,
        \end{equation}
where $m_{0}$ is a positive (by the dominant energy condition), dimensionless, constant of integration. Clearly, if $a$ increases indefinitely with time, then $\rho$ will tend to zero.

On using these expressions into the Einstein's field equation, we end up with the following evolution equation for $h$
        \begin{equation}\label{EOM:h}
         \tau_{0}\frac{\dd{h}}{\dd{t}} = 1 - h^{2} - \frac{2\kS^{2}}{9} - \frac{(1 + 3\wm)m_{0}}{6a^{3(1 + \wm)}}~,
        \end{equation}
which is nothing but the Raychaudhuri equation, as well as the following evolution equation for $\kS$
        \begin{equation}\label{EOM:kS}
         \tau_{0}\frac{\dd\kS}{\dd{t}} = 3 - 3h\kS - 3h^{2} + \frac{\kS^{2}}{3} + \frac{m_{0}}{a^{3(1 + \wm)}}~.
        \end{equation}
These evolution equations, along with~\eqref{EOM:a} and~\eqref{EOM:aS}, and the `initial value constraint equation' (\ie, the `time-time component' of the Einstein's equations)
        \begin{equation}\label{EOM:HC}
         h^{2} = \frac{\kS^{2}}{9} + 1 - \frac{\tau_{0}^{2}}{3a_{\gamma}^{2}} + \frac{m_{0}}{3a^{3(1 + \wm)}}~,
       \end{equation}
form a complete set of first order ordinary differential equations which determines the evolution of any appropriate initial data set (\ie, suitable values of the four functions $a$, $\aS$, $h$ and $\kS$ at some `initial moment').

We may also note parenthetically, that the above equations receive only minimal modifications if we let go of the assumption about the equation of state~\eqref{KaSa:eos}. In that case, every occurrence of $m_{0}a^{-3(1 + \wm)}$ in equations~\eqref{EOM:h}-\eqref{EOM:HC} should be replaced by a $\rho\tau_{0}^{2}$, while every occurrence of $\wm\rho$ should be replaced by a $p$. The equations may then represent more general situations involving multiple kinds of matter including the presence of dynamical fields (\eg, scalar field) etc. These more general versions of the equations are enough to obtain the results presented in section~\ref{section:KSKS} below.

Two exact solutions of the above sets of equations are particularly relevant for the discussion of de Sitterization (see, \eg,~\cite{Goliath:1998na} and references therein for further details on these well known solutions). They both arise when the matter terms in the above equations are set to zero. The first class of solutions are given by the following expressions for the metric functions
        \begin{equation}\label{soln:dSS}
         a_{z} = \sqrt{\frac{\tau^{2}}{\tau_{0}^{2}} - \cg + \frac{2\mu\tau_{0}}{\tau}}~, \qquad\qquad a_{\gamma} = \tau~,
        \end{equation}
where $\tau$ is a time function defined through the relation
        \begin{equation}\label{def:tau}
         \frac{\dd\tau}{\dd{t}} = a_{z}~,
        \end{equation}
and the parameter $\mu$ arises as a constant of integration. In fact, it can be shown (\eg, along the lines of~\cite{Bardeen:1973gs}) that the constant $\mu$ is essentially the `conserved charge' associated with the $\partial_{z}$ Killing symmetry. When $\mu$ is set to zero, these solutions represent various covers of parts of the global de Sitter manifold (for the different allowed values of $\cg$). Likewise, the solutions for non-zero $\mu$ cover patches of the de Sitter-Schwarzschild spacetimes. However, the real relevance of these solutions in the context of de Sitterization stems from the fact that these solutions~\emph{approximately} describe a de Sitterizing solution at `late times' (\ie, as $t \to \infty$), much like how the Schwarzschild solution approximately describes the `far away' region of a static, spherically symmetric and asymptotically flat spacetime. In other words,~\emph{the `departure' of a de Sitterizing solution with respect to that in~\eqref{soln:dSS} becomes increasingly smaller as time grows}.

The second exact solution that we wish to discuss exists only when $\cg = 1$, \ie, for the proper Kantowski-Sachs spacetimes. This solution is given by
        \begin{equation}\label{soln:dS2XS2}
         \begin{split}
          a_{z} & =
           \begin{cases}
                          \cosh(\sqrt{3}\,t/\tau_{0}), \qquad & |h| < 1/\sqrt{3}~, \\
                       \exp(\pm\sqrt{3}\,t/\tau_{0})~, \qquad & |h| = 1/\sqrt{3}~, \\
            \text{sign}(t)\sinh(\sqrt{3}\,t/\tau_{0}), \qquad & |h| > 1/\sqrt{3}~,
           \end{cases} \\
          & \\
          a_{\gamma} & = \frac{\tau_{0}}{\sqrt{3}}~.
         \end{split}
        \end{equation}
The following features of the solution are noteworthy:
        \begin{itemize}
         \item The solution consists of five disconnected branches, namely the parts with $h < -1/\sqrt{3}$ and $h > 1/\sqrt{3}$, the two fixed points of the equations of motion for $h = \pm(1/\sqrt{3})$, and the branch with $|h| < 1/\sqrt{3}$. The two fixed points are the only solutions for which $h$ is constant at all times.
         \item The branch for $h > 1/\sqrt{3}$ occurs for $t > 0$; in fact $a_{z} \to 0$ and $h$ diverges to $+\infty$ as $t \to 0^{+}$, and the solution cannot be extended past $t = 0$. As $t \to +\infty$, $h$ asymptotes to the limiting values of $1/\sqrt{3}$.
         \item The branch for $h < -(1/\sqrt{3})$ describes the `time reversed' scenario of the above. It exists only for $t < 0$, $a_{z} \to 0$ and $h$ diverges to $-\infty$ as $t \to 0^{-}$ and the solution cannot be extended past $t = 0$.
        \end{itemize}
Clearly, this solution does not describe a de Sitterizing solution. Indeed, the existence of this solution is ultimately the reason why the $\cg = 1$ Kantowski-Sachs family of solutions do not necessarily de Sitterize; generic initial states may also either recollapse or, with very finely tuned initial conditions, may approach the above fixed points (see \eg,~\cite{Goliath:1998na} or~\cite{Coley:2003mj}) for a more detailed discussion). 

A relevant question to ask here is whether there exists any condition on initial states of the $\cg = 1$ models which guarantee that they~\emph{do} de Sitterize. We will try to present our conclusion here through a semi-quantitative analysis, backed up by some heuristic arguments. The following arguments are modeled after the discussion of the Bianchi IX case in~\cite{Wald:1983ky}, and these results can be fully corroborated through more careful analysis of the equations of motion (an outline of which is presented in section~\ref{section:KSKS:soln}).

Now, unlike the $\cg \leqq 0$ cases, we cannot immediately apply Wald's arguments to the the $\cg = 1$ case. However, we may still argue that if an initially expanding $\cg = 1$ solution also satisfies the following condition at an instant
        \begin{equation}\label{KaSa:crit}
         \Lambda \geqq \frac{\threeR}{2} \qquad\qquad\iff\qquad\qquad a_{\gamma} \geqq \frac{\tau_{0}}{\sqrt{3}}~,
        \end{equation}
then the condition must continue to hold and such a solution must de Sitterize. Note that the exact solution~\eqref{soln:dS2XS2} saturates the lower bound.

To arrive at the above conclusion, we note that if an initially expanding solution (\ie, one which satisfies $h > 0$) also satisfies the condition~\eqref{KaSa:crit} as well as the dominant energy condition (\ie, $m_{0} > 0$), then the initial value constraint equation~\eqref{EOM:HC} implies $h \geqq \frac{1}{3}\kS$. Hence, by the following identity
        \begin{equation}\label{KaSa:dcrit}
         \tau_{0}\frac{\dd}{\dd{t}}\left(\Lambda - \frac{\threeR}{2}\right) = \frac{2}{a_{\gamma}^{2}}\left(h - \frac{\kS}{3}\right)~,
        \end{equation}
which holds as an easy consequence of the relations~\eqref{KaSa:3R}-\eqref{EOM:aS}, we may conclude that the condition~\eqref{KaSa:crit} continues to holds, and in fact, even more strongly as time goes on. This in turn keeps $h > 0$ (in fact, $h$ can never be zero by the constraint equation) so that the solution is ever expanding. That would then also mean that both the `three-curvature' and the matter terms in the initial value constraint equation must vanish asymptotically. Therefore, assuming that $\kS$ also vanishes asymptotically, the constraint equation would force $h$ to approach unity in the same limit, indicating de Sitterization.

In fact, we may justify the asymptotic vanishing of $\kS$ as follows: based on the behavior of the exact de Sitter solutions expressed  in the form of the metric~\eqref{KaSa:met}, we expect the scale factor to diverges asymptotically as $a \sim (\tau/\tau_{0})$ for de Sitterizing solutions, where $\tau$ is the time function defined through the relation~\eqref{def:tau}. Now, the equation of motion for $\kS$~\eqref{EOM:kS} implies the following~\emph{exact} relation
        \begin{equation}\label{Smarr:kS}
         a^{3}\kS = \frac{\cg\tau}{\tau_{0}} + \text{constant}~,
        \end{equation}
where the `constant' piece is a constant of integration. This relation then explicitly shows that for a de Sitterizing solution $\kS$ must vanish asymptotically at least as $a^{-2}$ or faster. Admittedly our arguments are rather heuristic, but as already mentioned, they can be properly justified through a careful analysis of the equations of motion.

Clearly, every de Sitterizing Kantowski-Sachs solution must satisfy the strict inequality~\eqref{KaSa:crit} at least at one instant during its evolution (and hence at all times subsequently), since $\threeR$ must vanish asymptotically. In other words, the strict inequality part of condition~\eqref{KaSa:crit} is necessary and sufficient for de Sitterization. This also shows that any solution which does~\emph{not} de Sitterize must strictly violate the bound~\eqref{KaSa:crit}. However, a complete understanding of the entire solution space can be obtained through a proper analysis within the framework of dynamical systems. To further appreciate how the behavior of solutions of Kantowski-Sachs models are affected by the presence of a positive cosmological constant, see~\cite{Collins:1977fg} and~\cite{Goliath:1998na}.

So far, we have reviewed the essential details about the Kantowski-Sachs family of solutions that will be relevant for understanding de Sitterization. In the following section, we will look at such de Sitterizing solutions in terms of their generalized Kerr-Schild representations.
%======================================================================================================================
\section{Generalized Kerr-Schild representations}\label{section:KSKS}
A straightforward computation of the conformal (Weyl) curvature of the metric~\eqref{KaSa:met} reveals that it is of Petrov type D. Such algebraically special metrics often admit (generalized) Kerr-Schild representations of the form~\eqref{KerrSchild:met}\footnote{Research into (generalized) Kerr-Schild representations has a rather long history, and is itself a pretty mature subject. A good entry point to this topic is the book~\cite{Stephani:2003tm}; see especially chapter 32. While we have obtained the results in equations~\eqref{soln:Omega:gen} and~\eqref{soln:Phi:gen} independently (following the route outlined in the main text) it is quite likely that these results have already appeared in the literature in the past; unfortunately we haven't been able to locate such a source. It also seems highly plausible that one may be able to modify the results of reference~\cite{1987CQGra...4.1449S} to derive ours, but we have not verified this.}. As the first step towards illustrating our main proposal, we will demonstrate this to hold for the Kantowski-Sachs family of solutions.

To that end, we proceed as follows. Let $\met_{a b}$ be a metric of the Kantowski-Sachs family~\eqref{KaSa:met}. It is then easy to verify that the one-form $\ell = -\frac{1}{\sqrt{2}}(a_{z}^{-1}\dd{t} + \dd{z})$, where we have suppressed all tensor indices, is one of the two repeated principal null one-form of the Weyl curvature tensor. We want to express $\met_{a b}$ in the generalized Kerr-Schild form~\eqref{KerrSchild:met} using the above mentioned null one-form (this is just a choice; we could have used the other principal null one-form as well). Furthermore, as already discussed in our introductory remarks, we want the functions $\Omega$ and $\Phi$ to respect all the Killing symmetries of $\met_{a b}$. Since this property is also satisfied by the null one-form $\ell_{a}$ (being a principal null vector of the Weyl curvature of $\met_{a b}$), the metric $\tilde{\met}_{a b}$ also respects all the Killing symmetries of the physical metric. Finally, we also require $\tilde{\met}_{a b}$ to be a conformally flat Einstein metric satisfying $\tilde{R}_{a b} = \tilde{\Lambda}\tilde{\met}_{a b}$, where $\tilde{R}_{a b}$ is the Ricci curvature tensor of $\tilde{\met}_{a b}$ and $\tilde{\Lambda}$ is a constant which we keep different from the physical cosmological constant $\Lambda$ for the moment (this is for book-keeping purposes; the constancy of $\tilde{\Lambda}$ follows, of course, from the `twiddled version' of the contracted Bianchi identity satisfied by $\tilde{\met}_{a b}$). Since $\tilde{\met}_{a b}$ is a conformally flat Einstein metric satisfying all symmetries of ${\met}_{a b}$, it is fairly easy to show that $\tilde{\met}_{a b}$ is a maximally symmetric metric whose curvature is dictated by $\tilde{\Lambda}$.

To determine, now, the functions $\Omega$ and $\Phi$ we simply need to relate the Ricci curvatures of the two metrics (see, \eg,~\cite{Wald:1984rg}) and impose the equations of motion~\eqref{EOM:a},~\eqref{EOM:aS},~\eqref{EOM:h}-\eqref{EOM:HC}. We wish to stress here that our results in this section do~\emph{not} depend on whether or not the matter follows an equation of state of the form~\eqref{KaSa:eos}. Rather, they hold as long as the matter flow lines are orthogonal to the homogeneous hypersurfaces and the matter stress tensor is describable in terms of an energy density $\rho$ and a pressure $p$ (but no anisotropy term is allowed in the stress tensor).

The aforementioned operations then yield relations involving $\Omega$ and $\Phi$ which can be solved to determine these functions. It turns out that these relations are sensitive to whether or not the quantity $h - \frac{1}{3}\kS$ vanishes. However, the only solution which satisfies the condition $h - \frac{1}{3}\kS = 0$ everywhere is the exact solution~\eqref{soln:dS2XS2}. However. since this is also a non-de Sitterizing solution, we will henceforth assume
        \begin{equation}
         h - \frac{1}{3}\kS \neq 0~,
        \end{equation} 
without sacrificing any generality. Under such conditions, $\Omega$ can be shown to satisfy the following second order~\emph{linear} differential equation
        \begin{equation}\label{EOM:Omega:gen}
         \frac{\dd^{2}\Omega}{\dd{t}^{2}} - \frac{1}{\tau_{0}}\left(h + \frac{2}{3}\kS\right)\frac{\dd\Omega}{\dd{t}} + \frac{(\rho + p)}{2}\Omega = 0~.
        \end{equation}
Quite remarkably, the most general solution of the above equation can be presented in the following compact form
        \begin{equation}\label{soln:Omega:gen}
         \Omega = \frac{a_{\gamma}}{\tau_{0}}(c_{1} + c_{2}F)~,
        \end{equation}
where the (dimensionless) function $F$ is defined through the equation
        \begin{equation}\label{def:F}
         \frac{\dd{F}}{\dd{t}} = -\frac{\tau_{0}a_{z}}{a_{\gamma}^{2}}~,
        \end{equation}
while $c_{1}$ and $c_{2}$ are (dimensionless) constants of integration. Note that if $\Omega$ is a solution of~\eqref{EOM:Omega:gen} then so is any constant times $\Omega$. This is ultimately due to the fact that the initial value of $\Omega$ is physically irrelevant, as it represents a constant conformal transformation of $\tilde{\met}_{a b}$ and hence can be factored out. This redundancy will be exploited below. We should also note that the function $F$ is only defined above up to a constant (of integration). We will fix this constant by requiring that $F$ vanishes as one approaches the de Sitter fixed point. 

The above analysis also yields a~\emph{linear algebraic} equation for $\Phi$, whose solution is 
        \begin{equation}\label{soln:Phi:gen}
         \Phi = -a_{z}^{2} + \frac{1}{c_{2}^{2}}\left[\frac{\tilde{\Lambda}a_{\gamma}^2}{3} - \cg\Omega^{2}\right]~.
        \end{equation}
This, along with the general solution for $\Omega$ above, constitutes the complete solution of the Kerr-Schild representation problem. Stated differently, we have shown that with the sole exception of the exact solution~\eqref{soln:dS2XS2},~\emph{any} solution of the equations~\eqref{EOM:a},~\eqref{EOM:aS},~\eqref{EOM:h}-\eqref{EOM:HC} can be expressed in a generalized Kerr-Schild form~\eqref{KerrSchild:met} in terms of an $\Omega$ given as in~\eqref{soln:Omega:gen} and a $\Phi$ given as in~\eqref{soln:Phi:gen}. We emphasize once more that our derivations do not assume any relationship between the matter's pressure and energy density, let alone an equation of state of the form~\eqref{KaSa:eos}. Therefore, the above results are applicable to a much broader class of matter than perfect fluids (as long as there is no anisotropy term in the matter stress tensor).

Of course, our main interest lies in solutions which exhibit de Sitterization. To find their appropriate Kerr-Schild representations, we need to set $\tilde{\Lambda} = \Lambda$ so that $\Omega$ and $\Phi$ depend only on the physical cosmological constant, as well as make the choice $c_{1} = 0$ which is required to attain the correct asymptotic behavior of these functions. Furthermore, the freedom to scale $\Omega$ by a constant can be exploited to set $c_{2} = 1$. With these choices, the expression for $\Omega$ reduces to
        \begin{equation}\label{soln:Omega}
         \Omega = \frac{a_{\gamma}F}{\tau_{0}}~,
        \end{equation}
while that for $\Phi$ becomes
        \begin{equation}\label{soln:Phi}
         \Phi = \frac{a_{\gamma}^{2}}{\tau_{0}^{2}} - a_{z}^{2} - \cg\Omega^{2}~.
        \end{equation}
Our final task, then, is to demonstrate that these functions indeed posses the correct asymptotic behavior as required off de Sitterizing solutions.

To achieve that goal, we need to have some description of the de Sitterizing solutions and subsequently evaluate $\Omega$ and $\Phi$ on them. This becomes a trivial exercise as far as the class of exact solutions~\eqref{soln:dSS} is concerned, and for them one ends up with
        \begin{equation}\label{Omega-Phi:dSS}
         \Omega = 1~, \qquad\qquad \Phi = -\frac{2\mu\tau_{0}}{\tau}~.
        \end{equation}
These expressions clearly show that for the exact solutions, the above functions do indeed have the right behavior about the de Sitter fixed point, as expected.

For more general de Sitterizing solutions with matter, the asymptotic behavior of the Hubble dictates that the function $F$, as defined in~\eqref{def:F}, must go as $a^{-1}$ asymptotically. This sets the asymptotic limit for $\Omega$ to one, according to~\eqref{soln:Omega}. Unfortunately, for the function $\Phi$, this exercise becomes a little more complicated, since it is not immediately obvious that the expression on the right hand side of~\eqref{soln:Phi} must vanish asymptotically for~\emph{all} kinds of matter consistent with our choices. Rather, we need to look at explicit properties of solutions to see this happen. However, such solutions can only be constructed either through some approximation technique, or numerically. We will explore the former possibility in the next section.
%======================================================================================================================
\section{Generalized Kerr-Schild representations of de Sitterizing solutions}\label{section:KSKS:soln}
Our plan, in this section, is to construct suitable asymptotic expansions of de Sitterizing solutions of the Kantowski-Sachs family about the de Sitter fixed point. To be able to do that, we need to make explicit choices for the matter's energy density and pressure. We thus adopt the perfect fluid model here and henceforth, and assume an equation of state of the form~\eqref{KaSa:eos}.

Since the solutions in the neighborhood of interest are all expanding, we may be inclined to use the scale factor as a possible time function and represent everything else as functions of it. This idea is actually lucrative given that the matter terms in equations~\eqref{EOM:h}-\eqref{EOM:HC} are explicit functions of the scale factor. Unfortunately, it is easy to show that a series analysis in inverse integral powers of $a$ is not consistent unless $\wm$ is a multiple of $\frac{1}{3}$ (although, for all physically interesting kinds of fluid matter, including dust and radiation, $\wm$~\emph{is} a multiple of $\frac{1}{3}$). Instead, if we are willing to restrict ourselves only to~\emph{rational} values of $\wm$ (subject to the bounds~\eqref{wm:SEC+DEC} as dictated by the assumed energy conditions), then we can use the quantity $\tilde{a}$ defined as
        \begin{equation}\label{def:tla}
         \tilde{a} = a^{1/\nu}~, \qquad\qquad \nu = \frac{n - 1}{3(1 + \wm)}~,
        \end{equation}
as a viable alternative, where $n \geqq 3$ is the smallest integer such that the exponent $\nu \geqq 1$ is also an integer. Clearly, $\tilde{a}$ is a monotonic function of the scale factor and is therefore qualified to serve as a time function in a neighborhood of the de Sitter fixed point. We should also note that when (and only when) $\wm$ is a multiple of $\frac{1}{3}$, we have $\tilde{a} = a$.

We can now rewrite equations~\eqref{EOM:aS},~\eqref{EOM:h} and~\eqref{EOM:kS} with $\tilde{a}$ as the time function such that they read
        \begin{equation}\label{EOM:tla}
         \begin{split}
          \frac{\dd\aS}{\dd\tilde{a}} & = \frac{\nu\kS\aS}{h\tilde{a}}~, \\
          \frac{\dd{h}}{\dd\tilde{a}} & = \frac{\nu}{h\tilde{a}}\left[1 - h^{2} - \frac{2\kS^{2}}{9}\right] - \frac{(1 + 3\wm)\nu{m}_{0}}{6h\tilde{a}^{n}}~, \\
          \frac{\dd\kS}{\dd\tilde{a}} & = \frac{3\nu}{h\tilde{a}}\left[1 - h\kS - h^{2} + \frac{\kS^{2}}{9}\right] + \frac{\nu{m}_{0}}{h\tilde{a}^{n}}~.
         \end{split}
        \end{equation}
The initial value constraint equation~\eqref{EOM:HC}, on the other hand, should remain as is since there are no derivatives involved there. We then postulate series expansions for each one of the three functions $h$, $\aS$ and $\kS$ in inverse integral powers of $\tilde{a}$, and solve for the corresponding coefficients order by order. The required asymptotic behavior of each function can be achieved by appealing to the exact solutions~\eqref{soln:dSS} and demanding $\aS \to 1$, $h \to 1$ and $\kS \to 0$ as the de Sitter fixed point is approached. The upshot of the analysis can be summarized through the following expressions for these functions\footnote{The results in equation~\eqref{soln:hak(a)} have been verified using~\texttt{SageMath}, a free and open source computer algebra system.}
        \begin{equation}\label{soln:hak(a)}
         \begin{split}
            h & = 1 - \frac{\cg}{6a^{2}} + \frac{m_{0}}{6a^{3(1 + \wm)}} \\
              & \qquad\qquad + \frac{7\cg^{2} + m_{0}(2\cg - m_{0})\delta_{\wm,\,-\frac{1}{3}}}{72a^{4}} + \cdots~, \\
          \aS & = 1 - \frac{\cg}{2a^{2}} + \frac{\mu + \frac{1}{12}m_{0}\delta_{\wm,\,0}}{a^{3}} + \frac{\cg^{2}}{24a^{4}} + \cdots~, \\
          \kS & = \frac{\cg}{a^{2}} - \frac{3\mu + \frac{1}{4}m_{0}\delta_{\wm,\,0}}{a^{3}} \\
              & \qquad\qquad + \frac{\cg^{2} + \cg{m}_{0}\delta_{\wm,\,-\frac{1}{3}}}{6a^{4}} + \cdots~,
         \end{split}
        \end{equation}
which approximate each of them accurately up to $\order(a^{-4})$. Note that the above series makes mathematical sense only after replacing each occurrence of the scale factor $a$ with $\tilde{a}^{\nu}$, in accordance with~\eqref{def:tla}. This is especially true due to appearance of the $\order(a^{-3(1 + \wm)})$ term in the series for $h$ (\ie, when $\wm$ is not a multiple of $\frac{1}{3}$; note, however, that this term is irrelevant when $\wm > \frac{1}{3}$, since it contributes at an order which is higher than the accuracy of the series). That being said, the above expressions bring out the universal leading order behavior of the functions for all values of $\wm$ and $\cg$, and therefore are more informative this way. We should also point out that some of the coefficients in the above series are sensitive to the value of $\wm$ through their dependence on the Kronecker delta functions (which are non-zero only for the specified values of $\wm$).

These series thus represent de Sitterizing solutions of the Kantowski-Sachs family, for given values of $\wm$, $\cg$ and the constant of integration $\mu$ (the latter being associated with the Killing symmetry of $\partial_{z}$; recall the discussion in the paragraph following equation~\eqref{def:tau}), up to the specified accuracy.

In order to understand the dependence of the solutions on a more `conventional time function' (\eg, the function $\tau$ as defined in equation~\eqref{def:tau}), we may construct the time function itself as a series in $a$. In particular, for $\tau$ we may combine equations~\eqref{EOM:a} and~\eqref{def:tau} to form a differential equation for $\tau$ as a function of $a$ (and hence, of $\tilde{a}$). The resulting analysis yields
        \begin{equation}\label{soln:tau(a)}
         \begin{split}
          \frac{\tau}{\tau_{0}} = a & + \frac{\cg}{6a} - \frac{\mu + \frac{1}{12}m_{0}\delta_{\wm,\,0}}{3a^{2}} \\
                                    & + \frac{m_{0}}{6(2 + 3\wm)a^{2 + 3\wm}} \\
                                    & + \frac{\cg^{2} + \frac{1}{3}m_{0}(\frac{2}{3}\cg - m_{0})\delta_{\wm,\,-\frac{1}{3}}}{24a^{3}} + \cdots~,
         \end{split}
        \end{equation}
which is accurate up to $\order(a^{-3})$. In particular, the $\order(a^{-(2 + 3\wm)})$ term is again irrelevant when $\wm > \frac{1}{3}$. The above expression also explicitly shows $a \sim (\tau/\tau_{0})$ as expected; hence the metric functions $a_{z}$ and $a_{\gamma}$ also have similar asymptotic behavior.

Finally, we may determine the functions $\Omega$ and $\Phi$ by evaluating their expressions -- ~\eqref{soln:Omega} and~\eqref{soln:Phi}, respectively -- with the help of the series~\eqref{soln:hak(a)}. We thus end up with
        \begin{equation}\label{soln:Omega:series}
         \begin{split}
          \Omega = 1 & - \frac{m_{0}}{6(4 + 3\wm)a^{3(1 + \wm)}} \\
                     & - \frac{m_{0}(4\cg - 9m_{0})\delta_{\wm,\,-\frac{1}{3}}}{1080a^{4}} + \cdots~,
         \end{split}
        \end{equation}
which is accurate up to $\order(a^{-4})$ and
        \begin{equation}\label{soln:Phi:series}
         \Phi = -\frac{2\mu + \frac{1}{6}m_{0}\delta_{\wm,\,0}}{a} + \frac{\cg{m}_{0}\delta_{\wm,\,-\frac{1}{3}}}{9a^{2}} + \cdots~,
        \end{equation}
which is accurate up to $\order(a^{-2})$. We should also note that the $\order(a^{-3(1 + \wm)})$ term in $\Omega$ is irrelevant when $\wm > \frac{1}{3}$. The above asymptotic expansions clearly demonstrate that $\Omega \to 1$ and $\Phi \to 0$ as the de Sitter fixed point is approached. As a non-trivial check of our results, we also note that in the limit $m_{0} \to 0$, the above expressions approach their counterparts in equation~\eqref{Omega-Phi:dSS}. 

We have thus established, as promised, that the generalized Kerr-Schild representations~\eqref{KerrSchild:met} of de Sitterizing solutions of the Kantowski-Sachs family of spacetimes accurately capture the de Sitterization process. Further discussion of our results and their ramifications will be taken up in our concluding remarks in the following section.
%======================================================================================================================
\section{Summary and discussions}\label{section:conclusion}
Homogeneous cosmological models in the presence of a positive cosmological constant (and subject to some additional minor restrictions) are known to de Sitterize, \ie, evolve towards the de Sitter spacetime. In this paper we have proposed a procedure to demonstrate the evolution of a subset of such homogeneous models which admit genralized Kerr-Schild represenations. Our overall procedure can be summarized through the follows steps:
    \begin{itemize}
     \item Start with a generalized Kerr-Schild ansatz of the form~\eqref{KerrSchild:met}, \ie
      \begin{equation*}
       \met_{a b} = \Omega^{2}\tilde{\met}_{a b} - 2\Phi\ell_{a}\ell_{b}~,
      \end{equation*}
where $\met_{a b}$ is the metric on the homogeneous cosmological model of interest (\ie, the physical metric), and $\tilde{\met}_{a b}$ is a conformally flat Einstein metric whose curvature is governed by the physical cosmological constant. Needless to say, we are assuming here that $\met_{a b}$ admits a generalized Kerr-Schild representation like above. In this work, we have chosen $\met_{a b}$ to be the metric on the Kantowski-Sachs family of spacetimes~\eqref{KaSa:met} for illustrating our proposal.
     \item Impose all the Killing symmetries of $\met_{a b}$ on the functions $\Omega$ and $\Phi$.
     \item Pick $\ell_{a}$ to be one of the principal null one-forms of the conformal curvature of $\met_{a b}$ (that $\ell_{a}$ must be one such principal null one-form is dictated by general properties of such representations~\cite{Stephani:2003tm}). 
     \item Use the relation between the Ricci curvatures of the two metrics and the Einstein's equations to obtain equations for $\Omega$ and $\Phi$ in terms of the parameters of the physical metric and of the matter stress energy tensor.
     \item Analytically, numerically or perturbatively find solutions for $\Omega$ and $\Phi$.
    \end{itemize}

A generalized Kerr-Schild relation can be viewed as a transformation of one metric into another one. Typically, the goal is to represent a more `complicated' metric (say, the physical metric $\met_{a b}$) in terms of a `simpler' metric (\eg, the de Sitter metric $\tilde{\met}_{a b}$). However, such relations are purely mathematical in nature. In particular, the physical significance of the quantities that go into a representation like~\eqref{KerrSchild:met} depends on additional inputs and restrictions, \ie, they are not necessarily built into the representation itself.

The most general solutions for $\Omega$ and $\Phi$ that one could obtain at the end of the process summarized above offer an excellent illustration of this point. Even though they satisfy all the symmetry requirements, the general solutions do not automatically represent de Sitterization~\emph{unless} we make appropriate choices for the parameters that appear in the general solutions for $\Omega$ and $\Phi$. Said differently, if we were to use the most general solutions for $\Omega$ and $\Phi$ in the relation~\eqref{KerrSchild:met}, the metric $\tilde{\met}_{a b}$ would~\emph{not} have been the asymptotic limit of the physical metric. Such general solutions are usually helpful when Kerr-Schild representations are used for solution generating purpose (as they often are). Here, on the other hand, we are using such representations to describe the time evolution of the physical metric, and for that one has to `tune' the functions $\Omega$ and $\Phi$ appropriately. This `tuning' is achieved such that they represent de Sitterizing solutions, \ie, satisfy the asymptotic conditions $\Omega \to 1$ and $\Phi \to 0$. With the help of these `properly tuned' versions of these functions, it may be possible extend the domain of the Kerr-Schild representation `far away' from the de Sitter fixed point.

We have stressed that our results, and especially the expressions~\eqref{soln:Omega} and~\eqref{soln:Phi}, hold as long as the matter flow lines are orthogonal to the homogeneous hypersurfaces and the stress tensor does not include any anisotropy term. However, to establish de Sitterization conclusively we had to specialize to perfect fluid matter satisfying some simple equation of state, because without this specialization, we were unable to solve the equations of motion and evaluate the asymptotic behavior especially of the function $\Phi$. It will be interesting to study our proposal by considering other types of matter (including dynamical fields, \eg, scalar fields) and verify that our claims still hold.

It was also pointed out that we randomly picked one of the two repeated principal null one-forms of the Weyl curvature of the physical metric to construct the representation~\eqref{KerrSchild:met}. Hence, there is~\emph{not} a unique Kerr-Schild representation of the physical metric, since we can also construct a similar representation using the other principal null one-form. Such a relation would then give rise to a new pair of functions, say $\Omega'$ and $\Phi'$, which are analogous to but distinct from $\Omega$ and $\Phi$ respectively (although their respective asymptotic limits must agree). This new representation must describe the de Sitterizing process slightly differently, since the limits of the physical metric in the two representations are not equal to each other (even though in both cases the limits are maximally symmetric conformally flat Einstein metrics whose curvatures are determined by the physical cosmological constant). Hence, these representations describe slightly different ways to approach the asymptotic region of the spacetime. Consequently, the functions $\Omega'$ and $\Phi'$ must also contain valuable information about the asymptotics of de Sitterizing solutions. This makes it an interesting future endeavor to work out the expressions for these new functions, as well as figure out their precise relationships with the functions $\Omega$ and $\Phi$.

Our choice of using the Kantowski-Sachs family of spacetimes to illustrate our proposal was largely motivated by two factors. First of all, Kantowski-Sachs family of spacetimes are of Petrov type D (\ie, algebraically special) and are likely to admit Kerr-Schild representations like~\eqref{KerrSchild:met}. Secondly, other algebraically special homogeneous cosmological models which may potentially admit Kerr-Schild representations are all of Bianchi type, and therefore are guaranteed to de Sitterize by Wald's theorem~\cite{Wald:1983ky}. The proper Kantowski-Sachs models (\ie, the $\cg = 1$ cases of the Kantowki-Sachs family), on the other hand, do not respect Wald's theorem (since they are closed models, \ie, satisfy $\threeR > 0$), and hence are more interesting from the perspective of our proposal.

Of course, we need to test our proposal beyond the Kantowski-Sachs family of spacetimes to broaden its scope further and make it more useful. This offers a strong motivation to study other algebraically special homogeneous cosmological models along the lines presented in this paper. In particular, it will be interesting to explore all LRS Bianchi models in this approach. However, since all such models must necessarily de Sitterize, thanks to Wald's theorem, it seems very likely that our proposal will hold for any model which admits a generalized Kerr-Schild representation like~\eqref{KerrSchild:met}.

We wish to end our concluding remarks speculating on one possible application of our proposal. The relation~\eqref{KerrSchild:met}, as it stands, is not restricted by the assumption of homogeneity. Therefore, we may employ such a relation to study~\emph{homogenization} along with isotropization of an initially non-homogeneous and anisotropic cosmological spacetime (albeit of very special kinds). In fact, it is likely that special kinds Gowdy-type~\cite{scholarpedia:Gowdy} inhomogeneous and anisotropic models will fit the bill, and it will be interesting to derive properties of such models through the corresponding $\Omega$ and $\Phi$ functions. We reserve this and other possible applications of our proposal for future work.  
%======================================================================================================================
\section*{Acknowledgement}
KB would like to thank SERB CRG/2020/002035 for support. JB would like to thank the~\texttt{SageMath}~\cite{sage} and the~\texttt{SageManifolds}~\cite{Gourgoulhon:2014ywa} projects for developing and maintaining their free and open source computer algebra system (with extensive online support), which was very helpful for carrying out numerous calculations for the current project.
%======================================================================================================================
\appendix
%======================================================================================================================
\section{Outline of the proof of Wald's theorem}\label{appendix:WaldsTheoremProof}
Our goal in this appendix is to briefly summarize the main arguments in~\cite{Wald:1983ky}. As already noted in the main text, as long as the assumptions hold, we need not care either about the details of the matter present, nor about most of the dynamical equations governing the overall evolution of the system. Rather, we only rely on the `initial value constraint' equation and the Raychaudhuri equation to deduce the appropriate asymptotic behavior of the mean curvature (\ie, the trace of the extrinsic curvature) of the homogeneous hypersurfaces. In particular:
    \begin{itemize}
     \item the initial value constraint equation bounds the mean curvature from below by $\sqrt{3\Lambda}$ as long as the assumed conditions hold;
     \item the Rauchaudhuri equation shows that under the same conditions the evolution of the mean curvature is always bounded from above by a monotonically decreasing function which also asymptotes to the value $\sqrt{3\Lambda}$. \end{itemize}
These two conditions thus force the mean curvature to attain its asymptotic limiting value, namely $\sqrt{3\Lambda}$. Furthermore, the constraint equation and the energy conditions also imply that the trace-free part of the extrinsic curvature (a measure of anisotropy), as well as the matter stress tensor, both vanish in the same limit, leaving us with a void, homogeneous and isotropic spacetime permeated by a positive cosmological constant $\Lambda$. These are all characteristics of a locally de Sitter spacetime, and it only takes a little more arguing to establish that the spatial metric also approaches its required asymptotic form under the above conditions, thereby finalizing the proof. We note in passing that reference~\cite{Starobinsky:1982mr} arrives at similar conclusions, but under more specialized conditions and also following a very different route.
%======================================================================================================================
\section{de Sitterization of FLRW spacetimes}\label{appendix:FLRW}
It is natural to wonder whether there is an analogue of the relation~\eqref{KerrSchild:met} for FLRW spacetimes, especially for de Sitterizing solutions. Of course, FLRW spacetimes being conformally flat, a generalized Kerr-Schild type relation is impossible to hold. However, one could easily anticipate expressing a de Sitterizing FLRW metric $\met_{a b}$ as follows
	\begin{equation}
	 \met_{a b} = \Omega^{2}\tilde{\met}_{a b}~,
	\end{equation}
where $\tilde{\met}_{a b}$ is a conformally flat maximally symmetric Einstein metric whose curvature is governed by the physical cosmological constant $\Lambda$,~and $\Omega$ is the appropriate conformal factor which is expected to go to one asymptotically.~It then becomes pretty straightforward to verify that for a relation like the above to hold, $\Omega$ must satisfy a~\emph{linear} second order differential equation given by
        \begin{equation}
         \frac{\dd^{2}\Omega}{\dd{t}^{2}} - \frac{h}{\tau_{0}}\frac{\dd\Omega}{\dd{t}} + \frac{(\rho + p)}{2}\,\Omega = 0~,
        \end{equation}
where our notations, conventions and definitions of the various quantities involved run parallel to those presented in the main text. In particular, $\tau_{0}$ is the scale set by the cosmological constant according to~\eqref{def:tau0}, $h$ is the dimensionless Hubble parameter defined in terms of the scale factor $a$ as in equation~\eqref{EOM:a}, and $\rho$ and $p$ are the energy density and pressure of the matter which source the Einstein's equations (in addition to the cosmological constant). In fact, we only require the following equations of motion for our purpose
        \begin{equation}
         \frac{\dd{a}}{\dd{t}} = \frac{ha}{\tau_{0}}~, \qquad\qquad h^{2} = 1 - \frac{\cg}{a^{2}} + \frac{\rho\tau_{0}^{2}}{3}~,
        \end{equation}
the first one being the definition of the dimensionless Hubble parameter, the second one being one of the Friedmann equations (the initial value constraint equation, to be precise), and $\cg$ being a constant which is equal to $-1$, $0$ and $1$ for open, flat and closed FLRW models, respectively. The solutions for $\Omega$ which yield the correct asymptotic behavior are given by
        \begin{equation}
         \Omega =
          \begin{cases}
           a\sinh{F}~, \qquad & \cg = -1~, \\
           aF~,        \qquad & \cg = 0~, \\
           a\sin{F}~,  \qquad & \cg = 1~,
          \end{cases}
        \end{equation}
where the function $F$ is defined through the relation
        \begin{equation}
         \tau_{0}\frac{\dd{F}}{\dd{t}} = -\frac{1}{a}~.
        \end{equation}
The asymptotic behavior of the Hubble parameter for de Sitterizing FLRW solutions dictates that $F$ goes as $a^{-1}$ asymptotically, thereby demonstrating that $\Omega$ goes to unity as the de Sitter solution is approached.
%======================================================================================================================

%======================================================================================================================
\end{document}